\documentstyle[12pt]{article}
\newcommand {\beq}{\begin{equation}}
\newcommand {\eeq}{\end{equation}}
\newcommand {\beqa}{\begin{eqnarray}}
\newcommand {\eeqa}{\end{eqnarray}}
\newcommand {\beqan}{\begin{eqnarray*}}
\newcommand {\eeqan}{\end{eqnarray*}}
\newcommand {\n}{\nonumber \\}

\newcommand {\Romannumeral}[1]{\uppercase\expandafter{\romannumeral#1}}

\newcommand {\mrc} {\mbox{\scriptsize RC}}
\newcommand {\str} {\mbox{\scriptsize str}}
\newcommand {\ee}{\mbox{e}}

\newcommand {\dd}{\mbox{d}}

\newcommand {\del}{\partial}

\begin{document}
\setlength{\oddsidemargin}{0cm}
\setlength{\baselineskip}{7mm}  

\begin{titlepage}
 \renewcommand{\thefootnote}{\fnsymbol{footnote}}
    \begin{normalsize}
     \begin{flushright}
                 KEK-TH-437\\
                 December 1995
     \end{flushright}
    \end{normalsize}
    \begin{Large}
       \vspace{2cm}
       \begin{center}
         {\Large On the Definition of the Partition Function \\
in Quantum Regge Calculus} \\
       \end{center}
    \end{Large}

  \vspace{1cm}

\begin{center}
           Jun N{\sc ishimura}\footnote
           {E-mail address : nisimura@eken.phys.nagoya-u.ac.jp}\\
      \vspace{1cm}
        $\ast$ {\it Department of Physics, Nagoya University,}\\
               {\it Chikusa-ku, Nagoya 464-01, Japan}\\
      \vspace{2cm}


\end{center}
\hspace{5cm}

\begin{abstract}
\noindent
We argue that the definition of the partition function used recently
to demonstrate the failure of Regge calculus is wrong.
In fact, in the one-dimensional case, we show that there is a more natural
definition, with which one can reproduce the correct results.
\end{abstract}

\end{titlepage}
\vfil\eject



Nowadays, due to the precise measurement of the gauge coupling constants
at LEP, it is usual to consider the three interactions which are
described quite well by the Standard Model to be unified into
a Grand Unified Theory at about the energy scale of $10^{16}$ GeV.
On the other hand, it is well known that the gravitational interaction
between particles becomes nonnegligible at the Planck scale, which is
around $10^{19}$ GeV.
These two energy scales are remarkably close to each other
considering the ambiguity involved in the derivation of the above values.
The most natural interpretation of this fact is
that, around such a high energy scale, all the
four interactions, including gravity, are unified.
There are two possibilities for such unified theories at present.
One is string theory and the other is a unification
within ordinary field theory including gravity.

The problem we encounter when we try to study quantum gravity within
ordinary field theory in four dimensions is
that we cannot renormalize it perturbatively.
We therefore need to use some kind of nonperturbative approach.
In the path integral formalism, the quantization of the
geometry is performed by integrating over the metric field.
In order to make it well-defined, we need to regularize the theory.
There are two types of lattice regularization of quantum gravity.
One is dynamical triangulation and the other is Regge calculus.
In lattice regularization, the general coordinate invariance is not
manifest and whether it is restored in the continuum limit is a crucial
problem.
Dynamical triangulation is exactly solved in two dimensions
\cite{DT} and its
continuum limit is shown to reproduce Liouville theory
\cite{DDK}.
There is also a handwaving argument for the restoration of general coordinate
invariance in the continuum limit of dynamical triangulation
\cite{handwaving}.
{}From the viewpoint of numerical simulation, however,
dynamical triangulation is much harder than ordinary statistical systems
since we have to change the lattice structure dynamically,
which makes it difficult for us to write a vectorized code.
Numerical simulation of four--dimensional dynamical triangulation
has been performed now for three years, and
although it is promising that a second order phase transition
has been found by sweeping the gravitational constant in the
Einstein--Hilbert action \cite{4DQG},
it is still unclear whether we can take a sensible
continuum limit at the critical point.

Obviously we need a larger lattice, and
in this respect it is worth while studying the other type of lattice
regularization of quantum gravity, namely Regge calculus \cite{Regge}.
In this formalism, the lattice structure is fixed and the fluctuation of
geometry is represented with the integration over the link lengths
within the constraint of triangle inequalities.
The lattice structure can be taken to be a regular lattice, for example.
Since this system is nothing but a kind of ordinary statistical system,
it is generally easy to write
an efficient code using vectorization.
However, the crucial problem of this formalism is that
there is no plausible argument for recovering the general coordinate
invariance
as with dynamical triangulation.
Also we have no guiding principle for choosing
the measure for the link--length integration.
One natural strategy is to study the two--dimensional case
and to see if it reproduces the known results of
two--dimensional quantum gravity \cite{GrossHamber,BockVink,HolmJanke,NO}.

One of the most fundamental quantities in two--dimensional
quantum gravity is the
partition function for a fixed total area $A$,
which is known to have the following large $A$ behavior.
\beq
Z(A) \sim A^{\gamma_{\str}-3} \ee^{\kappa A}
\eeq
Here, $\kappa$ is subject to renormalization from
the cosmological constant in the bare action,
whereas the $\gamma_{\str}$, which is called string susceptibility,
is a universal quantity in the sense that it cannot
be changed by changing the bare action.
The explicit form for the string susceptibility $\gamma_{\str}$
is known as the KPZ formula
\cite{KPZ}, which has been derived also from Liouville theory \cite{DDK}
in accordance with the known result from matrix models \cite{DT}.
A fundamental test of Regge calculus is, therefore,
to see if we can reproduce the string susceptibility
in two--dimensional quantum Regge calculus.

Such an attempt was made by Gross and Hamber~\cite{GrossHamber}
a few years ago and by two other groups~\cite{BockVink} more recently,
from which the conclusion seems to be that Regge calculus
fails to reproduce the desired string susceptibility.
In Ref. \cite{NO}, however,
we pointed out that the definition of the partition
function in Regge calculus is subtle and
that the one they used might be wrong.
In this paper, we study the one--dimensional case
in order to make our arguments concrete.
We first show that by using their definition of the partition function the
measure is almost uniquely determined
by the requirement that it should reproduce
the correct continuum result.
We then show that, with this special choice of measure,
one cannot reproduce
the correct Green's functions.
On the other hand, we show that there is a class of measure
which reproduces the correct Green's functions.
We also give a natural definition of the partition function with which one
can reproduce the correct result for the above class of measure.

{}~

Let us first consider one--dimensional quantum gravity
in the continuum theory.
We consider a line segment parametrized with the parameter $\tau$.
Let the ends correspond to $\tau=0$ and $\tau=1$ respectively.
On the line segment, we put $D$ copies of
the matter field $X^{\mu}(\tau)$
$(\mu=1,2,\cdots,D)$ and the metric
$g(\tau)$.
We consider the action
\beq
S= \int \dd \tau \sqrt{g(\tau)}
\left( \frac{1}{2} g^{-1}(\tau)
\frac{\del X^{\mu}(\tau)}{\del \tau}
\frac{\del X^{\mu}(\tau)}{\del \tau}
+ \frac{1}{2} M^2  X^{\mu}(\tau) X^{\mu}(\tau) \right),
\eeq
which is invariant under the reparametrization $\tau' = f(\tau)$.
In the following we set $M^2=0$ for simplicity.

The partition function can be defined as
\beq
Z(L,X^{\mu},Y^{\mu})
= \int {\cal D}g {\cal D}X^{\mu} \ee^{-S[g,X]}
\delta \left( \int_0^1 \dd \tau \sqrt{g(\tau)} - L \right) ,
\eeq
where the path integral is performed with the constraints
$X^{\mu}(0) = X^{\mu}$ and $X^{\mu}(1) = Y^{\mu}$.
The measure for the path integral is defined
through the norms
\beqa
|| \delta g ||^2 &=& \int \dd \tau \sqrt{g(\tau)} g^{-2}(\tau)
(\delta g (\tau))^2  \\
|| \delta X ||^2 &=& \int \dd \tau \sqrt{g(\tau)}
(\delta X (\tau))^2,
\eeqa
so that it is
reparametrization invariant.
The partition function can be expressed as
\beq
Z(L,X^{\mu},Y^{\mu})=
\frac{1}{(\sqrt{2 \pi L})^D}
    \exp \left\{ -\frac{(X^{\mu}-Y^{\mu})^2}{2L} \right\}.
\label{eq:partitioncont}
\eeq

Let us consider the corresponding quantity in Regge calculus.
We divide a line segment into $n$ pieces.
The metric degrees of freedom are represented by the link lengths
$l_i$ ($i=1,2,\cdots,n$) and the matter field $X^{\mu}_i$
is sited on each node ($i=0,1,\cdots,n$).
Let us consider the following quantity.
\beqa
&&  Z_{\mrc}[n,L,X^{\mu},Y^{\mu}] \n
&=& \int \prod_{i=1}^n \dd l_i \rho _n (l_i)
    \prod_{i=1}^{n-1} \dd X^{\mu}_i
    \exp \left\{ - \sum_{i=1}^n
    \frac{(X^{\mu}_{i}-X^{\mu}_{i-1})^2}{2 l_i} \right\}
    \delta(\sum_{i=1}^n l_i - L )  \n
&=& \prod_{i=1}^n \{ \dd l_i \rho_n (l_i) ( 2 \pi l_i )^{D/2} \}
    \delta( \sum_{i=1}^n l_i - L )
    \frac{1}{(\sqrt{2 \pi L})^D}
    \exp \left\{ -\frac{(X^{\mu}-Y^{\mu})^2}{2L} \right\},
\label{eq:rcpartition}
\eeqa
where $X^{\mu}_0 = X^{\mu}$ and $X^{\mu}_n = Y^{\mu}$.
One definition of the partition function in
terms of the above quantity is
\beq
Z[L,X^{\mu},Y^{\mu}] =
\lim _{n\rightarrow \infty}
Z_{\mrc}[n,L,X^{\mu},Y^{\mu}]
\cdot f(n) \ee ^{-g(n) L},
\label{eq:partnaive}
\eeq
where $f(n)$ represents the overall renormalization and $g(n)$
represents the renormalization of the cosmological term.
This is the definition of the partition function
that corresponds to the one adopted in Refs. \cite{GrossHamber,BockVink}.
In order that the partition function thus defined may
agree with the continuum result (\ref{eq:partitioncont}), the condition
\beq
\lim_{n\rightarrow \infty}
\int \prod_{i=1}^n \dd l_i \varphi_n(l_i)
\delta(\sum_{i=1}^n l_i - L) \cdot f(n)^{-1} \ee ^{g(n) L}=1
\label{eq:naive}
\eeq
should be satisfied,
where $\varphi_n(l)=\rho_n(l)(2\pi l)^{D/2}$.
Using a Laplace transformation one finds that $\varphi_n(l)$ should
have the following asymptotic behavior for large $n$.
\beq
\varphi_n(l) \sim f(n)^{1/n} \frac{1}{\Gamma(\frac{1}{n})}
\frac{1}{l^{1-\frac{1}{n}}}\ee ^{-g(n) l},
\label{eq:naivemeasure}
\eeq
where $f(n)$ and $g(n)$ can be taken arbitrarily.
We can take, for example,
\beq
\varphi_n(l) = \frac{1}{\Gamma(\frac{1}{n})}
\frac{1}{l^{1-\frac{1}{n}}}\ee ^{-\lambda l}.
\eeq
As is seen here, the large $n$ limit is, essentially,
unnecessary in reproducing the continuum result.

Let us see then if we can reproduce the Green's function
which is defined as
\beq
G^{(1)} (L,L',X^{\mu},Y^{\mu},Z^{\mu})
=\int {\cal D}g {\cal D}X^{\mu} \ee^{-S[g,X]}
\delta \left( \int_0^1 \dd \tau \sqrt{g(\tau)} - L \right).
\eeq
The path integral is performed with the constraints
$X^{\mu}(0) = X^{\mu}$ and $X^{\mu}(1) = Y^{\mu}$ and
$X^{\mu}(\xi) = Z^{\mu}$, where
$\int_0^{\xi} \sqrt{g(\tau)} \dd \tau = L'$.
The continuum result is given by
\beqa
&&G^{(1)} (L,L',X^{\mu},Y^{\mu},Z^{\mu}) \n
&=& Z(L',X^{\mu},Z^{\mu}) \cdot Z(L-L',Z^{\mu},Y^{\mu}) \n
&=&  \frac{1}{(\sqrt{2 \pi L'})^D}
    \exp \left\{ -\frac{(X^{\mu}-Z^{\mu})^2}{2L'} \right\}
    \frac{1}{(\sqrt{2 \pi (L-L')})^D}
    \exp \left\{ -\frac{(Z^{\mu}-Y^{\mu})^2}{2(L-L')} \right\}.
\label{eq:greencont}
\eeqa
One can check explicitly that by integrating over $Z^{\mu}$ and $L'$,
one gets $Z(L,X^{\mu},Y^{\mu})$.

Let us consider the following quantity in Regge calculus.
\beqa
&& G^{(1)}_{\mrc}[n,L,L',X^{\mu},Y^{\mu},Z^{\mu}] \n
&=& \int \prod_{i=1}^{n}  \dd l_i \rho_n(l_i)
  \sum_{n'=1}^{n-1}
 \int
    \prod_{i=1}^{n'-1} \dd X^{\mu}_i
    \prod_{i=n'+1}^{n-1} \dd X^{\mu}_i \n
&&    \exp \left\{ - \sum_{i=1}^n
    \frac{(X^{\mu}_{i}-X^{\mu}_{i-1})^2}{2 l_i} \right\} \cdot
     \delta \left( \sum_{i=1}^{n'} l_i - L' \right)
    \delta \left( \sum_{i=n'+1}^n l_i - (L-L') \right) \n
&=& \int \prod_{i=1}^n \{ \dd l_i \rho_n (l_i) ( 2 \pi l_i )^{D/2} \}
    \sum_{n'=1}^{n-1}
  \delta \left( \sum_{i=1}^{n'} l_i - L' \right)
  \delta \left( \sum_{i=n'+1}^n l_i - (L-L') \right) \n
&&    \frac{1}{(\sqrt{2 \pi L'})^D}
    \exp \left\{ -\frac{(X^{\mu}-Z^{\mu})^2}{2L'} \right\}
    \frac{1}{(\sqrt{2 \pi (L-L')})^D}
    \exp \left\{ -\frac{(Z^{\mu}-Y^{\mu})^2}{2(L-L')} \right\},
\eeqa
where $X^{\mu}_0 = X^{\mu}$, $X^{\mu}_n = Y^{\mu}$ and
$X^{\mu}_{n'} = Z^{\mu}$.
Note that, integrating over $Z^{\mu}$ and $L'$, one gets
$(n-1) Z_{\mrc}(n,L,X^{\mu},Y^{\mu})$.
Let us concentrate on the $L'$ dependence of
$G^{(1)}_{\mrc}[n,L,L',X^{\mu},Y^{\mu},Z^{\mu}]$,
which is not subject to
the overall renormalization or the renormalization of the
cosmological constant.
In order to get the correct $L'$ dependence of the continuum
result (\ref{eq:greencont}),
\beq
\int \prod_{i=1}^n \{ \dd l_i \varphi_n (l_i)  \}
    \sum_{n'=1}^{n-1}
  \delta \left( \sum_{i=1}^{n'} l_i - L' \right)
  \delta \left( \sum_{i=n'+1}^n l_i - (L-L') \right)
\label{eq:deltadelta}
\eeq
should be independent of $L'$.
Using the $\varphi_n(l)$ shown in eq. (\ref{eq:naivemeasure}),
the expression (\ref{eq:deltadelta}) reduces to
\beq
\frac{f(n) \ee^{-g(n)L}}{L'(L-L')}
\sum_{n'=1}^{n-1} \frac{1}{\Gamma(\frac{n'}{n})
\Gamma(1-\frac{n'}{n}) }
L'^{\frac{n'}{n}} (L-L')^{1-\frac{n'}{n}}.
\eeq
For sufficiently large $n$ the summation over $n'$ can be replaced with the
integral
\beqa
&\sim&  \frac{n f(n) \ee^{-g(n)L}}{L'(L-L')}
\int_0^1 \dd x  \frac{1}{\Gamma(x) \Gamma(1-x) }
L'^{x} (L-L')^{1-x} \n
&=& \frac{n f(n) \ee^{-g(n)L}}{L'(L-L')}
\frac{L}{\pi^2 + ( \ln \frac{L'}{L-L'})^2},
\eeqa
which means that we cannot reproduce the correct $L'$ dependence
for the Green's function (\ref{eq:greencont})
so long as we stick to reproducing the partition function
using the definition (\ref{eq:partnaive}).

Let us instead consider a class of measure which can be parametrized as
\beq
\varphi_{n} (l) = \frac{\lambda^N}{\Gamma(N)} l^{N-1}
\ee ^{-\lambda l},
\label{eq:Poissonmeasure}
\eeq
where $N$ and $\lambda$ generally depend on $n$.
When we take $N=1/n$, it reduces to the one considered above.
Eq. (\ref{eq:deltadelta}) then becomes
\beq
\frac{\lambda^{Nn} \ee^{-\lambda L}}{L'(L-L')}
\sum_{n'=1}^{n-1} \frac{1}{\Gamma(N n')
\Gamma(N(n-n')) }
L'^{N n'} (L-L')^{N(n-n')} .
\label{eq:generalN}
\eeq
When $Nn \gg 1$, using Stirling's formula,
the summation in the above expression is given by
\beq
\frac{Nn}{2\pi} \ee^{-Nn \ln Nn + Nn} \sum_{n'=1}^{n-1}
\sqrt{x(1-x)} \ee ^{-Nn f(x)},
\eeq
where $x=n'/n$ and
\beq
f(x) = x \ln \frac{x}{L'} + (1-x) \ln \frac{(1-x)}{(L-L')}.
\eeq
When $\frac{1}{\sqrt{Nn}} \gg \frac{1}{n}$, the summation can be evaluated
by integration in the large $n$ limit.
\beq
\sum_{n'=1}^{n-1} \rightarrow n \int_0^1 \dd x
\eeq
Using the saddle--point method, the expression (\ref{eq:generalN}) can be
evaluated and yields
\beqa
&&\frac{\lambda^{Nn} \ee^{-\lambda L}}{L'(L-L')} \cdot
\frac{Nn}{2\pi} \ee^{-Nn \ln Nn + Nn} n \cdot
\sqrt{\frac{L'}{L} \left( 1-\frac{L'}{L} \right )} L^{Nn}
\sqrt{\frac{2\pi}{Nn}\frac{L'}{L} \left( 1-\frac{L'}{L}\right)} \n
&\simeq& \frac{\lambda ^{Nn}}{\Gamma (Nn)} \ee^{-\lambda L} L^{Nn-1}
\frac{n}{L}.
\eeqa
Thus the $L'$ dependence disappears in the last expression, which means
that the correct Green's function can be reproduced as long as
\beq
 \frac{1}{n} \ll N \ll n
\label{eq:condition}
\eeq
is satisfied.

Let us then reconsider the partition function for this class of measure
(\ref{eq:Poissonmeasure}) with the above condition (\ref{eq:condition}).
Eq.(\ref{eq:rcpartition}) can be given by
\beq
 Z_{\mrc}[n,L,X^{\mu},Y^{\mu}]
=\frac{\lambda^{Nn}}{\Gamma (Nn)} L^{Nn-1} \ee^{-\lambda L} \cdot
Z(L,X^{\mu},Y^{\mu}).
\eeq
If we take the $\lambda \rightarrow \infty$ limit
so that $Nn/\lambda$ is kept to a constant,
say to $\bar{L}$, the prefactor of $Z(L,X^{\mu},Y^{\mu})$ in the above
expression tends to $\delta (L-\bar{L}) $.
Therefore one finds that
\beq
Z(\bar{L},X^{\mu},Y^{\mu})=
\lim_{n \rightarrow \infty} \int _0 ^{\infty} \dd L
Z_{\mrc} (n,L,X^{\mu},Y^{\mu}).
\label{eq:partcorrect}
\eeq
Similarly for the Green's function, we have
\beq
G^{(1)} (\bar{L},L',X^{\mu},Y^{\mu},Z^{\mu}) =
\lim_{n \rightarrow \infty} \frac{\bar{L}}{n} \int _0 ^{\infty} \dd L
G^{(1)}_{\mrc}[n,L,L',X^{\mu},Y^{\mu},Z^{\mu}] .
\eeq
One can easily generalize these results to
$n$--point Green's functions.

Thus we have found a class of measure
which reproduces not only the Green's functions
but also the partition function by employing a definition which
is different from the one used in previous works.
The problem of the definition (\ref{eq:partnaive}) is that
the $L$ dependence is extracted for a fixed $n$,
which means that
probing the $L$ dependence corresponds to looking at
configurations with different average link length.
On the other hand, the point of the definition (\ref{eq:partcorrect})
is that the actual total length of the line segment is not fixed externally
but that we are probing the $\bar{L}$ dependence, which corresponds to looking
at configurations with different $n$ with equal average link length.


\vspace{1cm}


To summarize, we find that, in the one--dimensional case,
employing the definition of the partition function
used recently to claim that Regge calculus fails to reproduce the
KPZ formula, we cannot reproduce both the partition function and the Green's
functions simultaneously.
By considering a class of measure parametrized as in
(\ref{eq:Poissonmeasure}) and the condition (\ref{eq:condition}),
that has to be satisfied
in order to reproduce the Green's functions correctly,
we find that there is
a definition of the partition function with which one can
reproduce the correct result.
We argue that this definition of the partition function
is more natural than the one used in the previous studies.
Thus the observations made in Refs. \cite{BockVink} are merely due to the
wrong definition of the partition function and they do not mean immediately
that
Regge calculus is wrong.
Also the fact that there is a successful measure at least for one--dimensional
quantum Regge calculus is quite encouraging.
Together with the observation made in Ref. \cite{NO} that
the loop length distribution for the ``baby loops'' has been
reproduced, we think that Regge calculus is still worth studying
by investigating other measures, for example,
before giving it up as a failure at this stage.

\vspace{1cm}

I would like to thank H. Kawai, T. Nakajima and N. Tsuda and
T. Yukawa for stimulating discussion.
I am also grateful to B. Hanlon for carefully reading the manuscript.

\end{document}